
\magnification 1200
\hoffset .25 in
\hsize  5 in
\centerline{{\tenbf Constraining the Right Handed Interactions from
Pion Condensate Matter}}
\vskip .5cm
\centerline{ Ashok Goyal and Sukanta Dutta}
\vskip .2 cm
\centerline{{\sl Department of Physics and Astrophysics, University of
Delhi, Delhi-110007, INDIA.}}
\centerline {{\sl and}}
\centerline {{\sl Inter-University Centre for Astronomy and Astrophysics,
Ganeshkhind, Pune,411007,INDIA.}}
\vskip 3cm
\centerline{{\bf Abstract}}
\vskip .5cm
\baselineskip=10 pt
\baselineskip =2\baselineskip
\indent We consider right handed neutrino emission from charged and neutral
pion condensate matter
likely to be present in the supernova core associated with SN 1987 A. This is
used
to constrain the strength of right handed interaction and we get excluded range
of values
for the right handed $W$ boson and extra neutral $Z'$ boson masses. For
vanishing $W_L-W_R$
mixing we obtain   $ (1.3-1.8)\,M_{W_L} \,\approx \leq\,M_{W_R} \, \leq\approx
100\, M_{W_L}
\, {\rm and}\, (1.3-1.8)\,M_{W_L}\, \approx \leq \, M_{N'}\,\leq\approx
\,225 \,M_{W_L}$
\vfil
\eject

\beginsection {1. Introduction}

\indent  The  microscopic composition of matter in the core of a neutron
  star  can  be  very different  from  normal nuclear matter. The
  conditions  in  the  core  might  be very close  to those where
  meson condensates $[1]$ could be formed, also there could be
an  abundance  of  quasifree  pions $[2]$ or a sufficiently large
concentration of protons  and hyperons $[3]$ to allow direct URCA
process  to  take  place. The matter in the core may even undergo
phase  transition  to its constitutent quark matter , giving rise
to  a  strange  matter  core  $[4].$ The exotic matter core has a
effect  of  modifying   the  kinematic  conditions  of reactions.
For  example  direct  $\beta-$  decay  is  inhibited  in  neutron
stars  essentially  because of simultaneous non conservation   of
energy    and    momentum    owing   to   Fermi-Dirac statistical
distribution.   The   presence of meson condensates or quasi free
pions  can  supply the  required  momentum  as  is  done  by  the
spectator nucleon in the case of modified URCA process, similarly
$\beta-$  decay of quarks and nucleons for large proton  fraction
in  the  core can take place. This would result in enhancement of
energy   loss  rate.  These exotic core compositions coupled with
the emission of weakly   interacting   exotic  particles  and  of
neutrinos   with  non-standard  properties  and  couplings  would
have  the consequence of potentially shortening the  duration  of
the  neutrino  burst  from  the core of a newly born, hot neutron
star  associated  with the early cooling phase of SN 1987 A. This
has  been  used to put  stringent  constraints  on the properties
and   couplings   of  exotic  particles  and   on   Dirac   mass,
anamolous   magnetic moment, right handed interactions {\it etc.}
of   neutrinos  $[5-9].$  Recently,  axion  emission  from  meson
condensate  matter  has   been   invoked  $[10]$  to  explain the
cooling  mechanisms  for  some  sources like PSR  $0833-45,$\quad
PSR  $0856+14,$\quad  $\gamma$ ray pulsar Gemingo and stars whose
observed  surface   temperatures are different from what could be
predicted in the standard cooling scenario.

In   this  paper,  we  assume the presence of charged and neutral
pion condensates in the core and   consider the emission of light
right  handed  neutrinos  {\it  via}  right  handed  charged  and
neutral  current  interactions  as in left-right symmetric models
of   weak   interactions  $[11]$ or  in some super-string inspired
models  $[12].$ Such  a  right  handed  neutrino  can be produced
in  the standard supernova core through  electron  capture,  pair
annihilation  or   neutrino  pair bremmstrahlung processes: \quad
$e^-    +  p  \longrightarrow  n  +  \nu_R,$\quad  $  e^+  +  e^-
\longrightarrow  \nu_R + \bar  \nu_R,$ \quad $n+ n \longrightarrow
n+n+\nu_R  + \bar \nu_R$ \quad and \quad $n +n \longrightarrow n +p
+e^-  + \bar \nu_R.$ \quad On account of high electron degeneracy
and  low proton fraction  the  first  two  processes are somewhat
suppressed.   In  the  presence  of  charged   and  neutral  pion
condensates,  we  have  the  following  reactions producing right
handed  neutrinos:
$$   n   +   \pi^-   \longrightarrow   n   +  e^-  +  \bar  \nu_R
\eqno{(1a)}$$
$$ n + \pi^0 \longrightarrow p + e^- + \bar \nu_R \eqno{(1b)}$$
$$  n   +   \pi^0   \longrightarrow   n   +   \nu_R  + \bar \nu_R
\eqno{(1c)}$$
\vskip .2 cm
\noindent   along with similar  processes  involving  protons  in
the  initial  state.  In  the presence of kaon  condensed  matter
analogous  reactions  involving kaons will also contribute to the
production   of  right  handed  neutrinos  but  will  be  Cabbibo
suppressed. Here we will not consider these processes.

In  section  2, we calculate the right handed neutrino emissivity
and  in  section  3,  we address the question of neutrino opacity
and  calculate  its mean free path in the presence of charged and
neutral pion condensates. The integrated luminosity is  then used
in  section  4  to  put  an  upper bound on the strength of right
handed  charged and neutral interactions. The lower bound is
obtained by the consideration of neutrino trapping in the core.

\beginsection {2. Calculation of Neutrino Emissivity}

\indent The   effective   weak  interaction  Hamiltonian  describing  the
processes  $(1)$  in the left-right  symmetric  model  $[12]$  is
given  by
$${\cal H_W}={G_R\over\sqrt 2}\,J_{R\mu}^h \, L_R^\mu \eqno{(2)}$$
 where $G_R^2 =B G^2_F$ with $ B= \xi^2  +  M^2_{W_L}/M^2_{W_R}$ for the
charged
current  and  $G^2_R  =  B'  G^2_F$  \quad with \quad
$B' = \xi^2 + M^2_{W_L} / M^2_{N'}$  for
the  neutral  current, $G_R$ is  the  strength  of
right  handed  current,  $\xi$ is the mixing parameter, $M_{W_R}$
is    the    mass    of    the   right  handed  $W$  boson  and
$M_{N'}$  is  the effective mass of right handed $Z$
boson, $J^h_{R\mu}$ and $L_R^\mu$ are the hadronic  and  leptonic
currents  respectively.  The  charged  hadronic  weak  current is
written as
$$   J^{hc}_{R\mu}  =V_\mu^1  +  A_\mu^1  +i \bigl(V_\mu^2
+A_\mu^2   \bigr)   \eqno{(3)}$$  \noindent  along  with  neutral
hadronic  weak  current as $$ J_{R\mu}^{h0} =  {1  \over  {\cos^2
\theta_W}}   \biggl[   V_\mu^3   +   A_\mu^3   \cos   2\theta_W -
J_\mu^{e.m.} \sin^2\theta_W \biggr] \eqno{(4)}$$

\beginsection {2.1 Charged pion Condensate}

\indent The   charged  pion  condensed  phase  is constructed by a chiral
rotation by the unitary operator $[13]$
$$U\bigl(  \pi^c;  \mu_\pi,  k_c,\theta \bigr) = exp \biggl(
i \int   d^3r\thinspace   \chi   V_3^0   \biggr) exp \Bigl(i
Q_1^5 \theta \Bigr) \eqno{(5)}$$
with $\chi =  k_c\cdot r  - \mu_\pi t$ acting on the
ground  state.  This  generates  the charged pion condensed phase
with  a  macroscopic  charged  pion field of  chemical  potential
$\mu_\pi,$ momentum  $k_c$  and chiral angle $\theta.$
The  participating   particles   for  $\nu_R$ emission
are the quasiparticles  $\eta$ and $\zeta$,  which are
superposition of the proton and neutron states:
$$\bigl\vert  \eta  (   p, \pm 1) \big\rangle = \cos\phi_c \bigr\vert
n ( p +k_c, \pm 1)\big\rangle\mp i \sin\phi_c\bigl \vert p(p+k_c,  \pm 1)
\big\rangle$$
$$\bigl\vert \zeta(  p, \pm 1)\big \rangle = \cos\phi_c\bigr \vert
p ( p - k_c, \pm  1)\big\rangle \pm i
\sin\phi_c \bigl\vert n(p - k_c, \pm 1)\big\rangle \eqno{(6)}$$
\vskip .2 cm
\noindent where $+ (-)$ refers  to  spin up  (down), and $\phi_c$ is
the  mixing  angle given by
$$ \phi_c = \biggl( {{g_Ak_c}  \over \mu_\pi }\biggr) \theta +O(\theta^2),$$
$g_A$  being  the   axial  vector  coupling  constant. The matrix
element squared and summed over spins is given  by
$$   \sum \bigl \vert M \bigr \vert^2 = H_{\mu \nu}\thinspace L^{\mu \nu}
\eqno{(7)}$$
\noindent with $$H_{\mu    \nu}   =   \sum_{s_1,s_2} \big\langle  \eta  (p_2,
s_2)\bigl \vert
\tilde J_\mu^{h}\bigr \vert   \eta(p_1,s_1)\big\rangle \big \langle \eta
(p_2,s_2)
\bigl \vert\tilde J_\nu^{h}\bigr\vert\eta(p_1,s_1)\big\rangle ^\dagger$$
$$L^{\mu \nu} = 8 \bigl( q_1^\mu q_2^\nu + q_1^\nu q_2^\mu
-q_1\cdot q_2   g^{\mu   \nu} -i\epsilon^{\mu\nu\alpha\beta}\, q_{1\alpha}
q_{2\beta}\bigr).$$
\vskip .2 cm
\noindent Here $ \tilde J_\mu^{h}$
\quad  is  the  rotated  hadronic  current  and  is  obtained  by
applying    chiral   rotation   $U$  of  equation  $(5)$  to
$J_\mu^h$  in order to include effect of charged pion condensate,
$p_1,p_2  $ are  the  four   momenta   of   incomming and
outgoing  quasiparticles  and  $q_1  ,  q_2$ are the four
momenta of leptons.

The   $\nu_R$   emission   process in charged pion condensate can
now be studied in terms  of quasiparticles. The process
$$ \eta \longrightarrow \eta + e^- + \nu_R \eqno{(8)}$$
\noindent  is   mediated   by   weak   flavor   changing  charged
current  given  in equation  $(3)$.  Applying the chiral rotation
$(5)$ we get
$$    \eqalignno{\tilde    J_{R\mu}^{hc}   &   = U \thinspace J_{R\mu}^{hc}
\thinspace U^{-1}\cr & = e^{i \chi} \biggl[ V^1_\mu +A^1_\mu
+  i \cos \theta \bigl(  V^2_\mu  +  A^2_\mu  \bigr)  + i
\sin \theta \bigl( V_\mu^3 +A_\mu^3 \bigr)  \biggr]  &(9)\cr}$$
\vskip .2 cm
\noindent In terms of this rotated current the square of the quasiparticle
$\beta  -$decay matrix element, summed over spins can be
obtained as
$$   \sum \bigl\vert  M \bigr \vert^2 = 2\,G^2_R \thinspace\theta^2 \biggl[ 1
+{ {g^2_A
k^2_c}  \over \mu_\pi^2 }\biggr] \Bigl[ \bigl( 1 + 3 g^2_A \bigr)
\omega_1 \omega_2 + \bigl(  1 -   g^2_A   \bigr)  \bar  q_1\cdot\bar
q_2   \Bigr]   \eqno   {(10)}$$
\vskip .2 cm
\noindent where   $  q_i
(\omega_i,\bar  q_i)$  correspond to the four momentum
of   $\nu_R$  and   electron   for   $i=1,2$ respectively.  The  emissivity
for  this  process  is  given  by
$$  \eqalignno  {\dot {\cal E}=& \prod _{i=1}^2 \Biggl[
\int  {{d^3p_i}  \over  {(2\pi)^3}}  \Biggr]  \prod  _{j=1}^2
\Biggl[   \int  { {d^3q_j}  \over  {2\omega_i  (2 \pi)^3}} \Biggr]
\bigl(2 \pi \bigr)^4  \delta^4 \bigl(  p_1 + k_c- p_2 - q_2 - q_1
\bigr) &\cr  & {\hskip 3cm }\omega_1  f \bigl( p_1 \bigr)
\Bigl[ 1- f\bigl( p_2
\bigr) \Bigr] \Bigl[ 1- f\bigl( q_2 \bigr) \Bigr] &(11)\cr }$$
\vskip .25 cm
\indent   Treating    the    nucleons   to  be  relativistic  and
nondegenerate  and  electrons to be degenerate,  all  phase space
integrals $(11)$ can be done analytically except one and  we  get
$$   \eqalignno  {\dot {\cal E}&= {G^2_R\, n_N\, \theta^2 \over (2
\pi^3)} \, T^6 \; \bigl(   1   +  3  g^2_A  \bigr)\; \biggl( 1 + {g^2_A
k_c^2  \over  \mu_\pi}  \biggr)  \;I \bigl(  {\mu_\pi
\over T} , {\mu_e \over T }\bigr)&\cr
&\simeq 1.6 \times 10^{34} B\, \theta^2\,  T^6_{10}\, \rho_{14}\,
\bigl(1  +  3  g^2_A  \bigr)&\cr
&{\hskip 1 cm}\times \biggl(  1 + {g^2_A k_c^2 \over  \mu^2_\pi}
\biggr) \Bigl( {938 MeV \over m_N^\ast} \Bigr)\, I\quad\quad {\rm ergs
\thinspace cm}^{-3}\thinspace {\rm s}^{-1}  & (12)\cr}$$
\vskip .2 cm
\noindent where  $n_N$ is the number density of nucleons in the core,
$\rho_{14}$ is the core density
in units of $10^{14}$
gms/cc,  $T_{10}$   is   the  temperature of the core in units of
10  MeV,  $m_N^\ast$  is  the  effective  mass of the
nucleon   and
$$I=  \int\displaylimits_0^\nu{ x^3 \bigl(\mu/T- x \bigr)^2
\over e^{( x - \nu )} + 1} \, dx $$
with $\nu=\bigl(\mu_\pi -\mu_e\bigr)/T.$
\vskip .25 cm
        The  other dominant process
$$\zeta \longrightarrow \eta + \nu_R + \bar {\nu_R}  \eqno{(13)}$$
\noindent  which   is  mediated  by the weak right handed neutral
current $(4)$ which under chiral rotation becomes
$$  \eqalignno { \tilde   J_{R\mu}^{h0}=&{1\over  \cos^2
\theta_W}\biggl[\bigl(V_\mu^3
+A_\mu^3 \cos2 \theta_W  \bigr)\cos\theta + \bigl(A_\mu^2 + V_\mu ^2 \cos
2  \theta_W  \bigr) \sin \theta &\cr & {\hskip 2 cm}-\,\sin^2\theta_W\,\bigl({1
\over 2}
V_\mu^Y +  V  _\mu  ^3  \cos \theta + A_\mu  ^3  \sin \theta\bigr)  \biggr]
&(14)\cr}$$
\vskip .2 cm
\noindent  The  relevant part of the  matrix  element squared and
summed over spins is given by
$$\sum  \bigl\vert  M   \bigr\vert^2 = 2\, G^2_R \, \biggl[ 3 g^2_A \theta^2 +
\biggl(  {g_A  k_c  \over   \mu_\pi}  \theta  +{  \cos  2\theta_W
\over    \cos^2  \theta_W}  \biggr)^2    \biggr]\,\omega_1   \*
\omega_2  \eqno{(15)}$$
\vskip .2 cm
\noindent  and  the emissivity  is  calculated  to  be
  $$   \eqalignno   {\dot  {\cal E} &= {G^2_R \over 240 \pi^3}\,
  n_N \,\mu_\pi^6\,\biggl[  3  g^2_A  \theta^2  +  \biggl( {g_A k_c \over
\mu_\pi}   \theta   +{   \cos   2\theta_W \over  \cos^2 \theta_W}
\biggr)^2  \biggr]&\cr
  &  \simeq  6.5 \times 10^{41}\, B'\,\rho_{14}\, \biggl( {938
MeV  \over  {m_N^\ast}}\biggr) \,\biggl( {\mu_\pi \over 250  MeV}
\biggr)^6  &\cr
& {\hskip .5 cm}\times \biggl[  3  g^2_A  \theta^2  +  \biggl(  {g_A k_c \over
\mu_\pi}   \theta   +{  \cos  2\theta_W  \over  \cos^2  \theta_W}
\biggr)^2 \biggr] \quad\quad{\rm ergs \,cm^{-3} \,s^{-1}} & (16)\cr}$$

\beginsection {2.2 Neutral Pion Condensate}

\indent Like  the  charged  pion  condensed phase, neutral pion condensed
phase  is  constructed  by  a chiral rotation of the ground state
with  the unitary operator $[10,14]$
$$U(\pi^0, \phi)= exp \bigl(i \int  d^3r\thinspace  A_3^0\thinspace  \phi
\bigr)  \eqno{(17)}$$
\noindent here $ \phi $ is the classical  pion  field  chosen  to
be  of  the  form   $\phi = A \sin k_0 z$  where  $A$
and   $k_0$   are   the  amplitude  and  momentum of the condensate
respectively.  In  this case the pion condensate forms a standing
wave  in  one  dimension  which  is  in $\hat Z$  direction. Following
reference  $[10]$,  the  wavefunction  for  nucleon quasiparticle
states are  given  by
$$ \eqalignno {\bigl\vert  \tilde  N  (p,  \pm  1)\big \rangle =\bigl
\vert &N (p, \pm 1)\big\rangle
\,\mp \,{A \kappa_0  \over  2} \; \Biggl[  {\bigl\vert  N  (p + k_0, \pm 1)
\big\rangle \over
\epsilon_N   (p   +k_0)  - \epsilon_N (p)}  &\cr
&\,+\,{\bigl\vert N (p - k_0),\pm 1
\big\rangle \over   \epsilon_N    (p   -   k_0)   -   \epsilon_N(p)}
\Biggr]\, +\,O(A^2)&(18)}$$
\vskip .2 cm
\noindent where $\epsilon_N$  is  the  single   particle   nucleon
energy,   and   $\kappa_0$  represents the modified pseudo-vertex
$[10]$.  The   hadronic  tensor  $ H^{\mu\nu} $ is
written as $$  H^{\mu\nu} = \sum_{s_1, s_2} \bigl \vert  B_{s_1,s_2}\bigl
\vert  ^2
H^{\mu\nu}_{s_1,s_2} \eqno {(19)}$$
where   $$   B_{s_1,s_2}   =   \int d^3r \,\phi_{s_2,p_2}^\ast (r)
\,\phi_{s_1,p_1} (r) \,e^{-i (q_1+q_2)\cdot r}  $$
and
$$  H^{\mu\nu}_{s_1,s_2}  = \big\langle\chi_{s_2}^N\bigl  \vert \tilde J^\mu_h
\bigr\vert \chi_{s_1}^N \big\rangle \, \big\langle \chi_{s_2}^N \bigl \vert
\tilde  J^\nu_h
\bigr\vert  \chi_{s_1}^N \big\rangle^\dagger .$$
\vskip .25 cm
\indent We now consider the  process
$$   \tilde  n  \longrightarrow \tilde p + e^- + \bar \nu_R \eqno
{(20)}$$
\noindent mediated by weak right handed flavor changing charged
current   which   under   rotation   by the unitary matrix $(17)$
becomes
$$ \tilde  J_{R\mu}^{hc} = e^{i \phi}\,\biggl[ V^1_\mu +A^1_\mu
+ i \bigl( V^2_\mu + A^2_\mu \bigr) \biggr] \eqno{(21)}$$
\noindent  Thus  the  hadronic current  remains  unchanged on the
chiral  rotation  and  the momentum is supplied  by  the  current
and  wavefunction both. The matrix element squared  and summed over spins
for the above process is given by
$$  \sum \bigl\vert M \bigr\vert^2 =   {4   G^2_R   g^2_A  (2  \pi^3)  A^2
\kappa_0^2   \over   \bigl(\omega_1  +\omega_2 \bigr)^2} \, \biggl[
\delta^3   (\bar   \beta  + \bar k_0 ) + \delta^3 (\bar  \beta  -
\bar   k_0   )  \biggr] \, \omega_1  \omega_2  \eqno{(22)} $$
\noindent where $ \bar\beta = \bigl( \bar p_1 -\bar p_2 -  \bar
q_2  -  \bar  q_1  \bigr).$ The emissivity now becomes
$$   \eqalignno   {\dot   {\cal E}  &= {G^2_R\, g_A^2\, \kappa_0^2
\,A^2 \over  4 \, \pi^3\, ( \mu_\pi - \mu_e  )^2}\,  n_N\, T^6\,
I\bigl( {\mu_\pi
\over T}, {\mu_e \over T }\bigr) &\cr
  &  \simeq  2.2  \times  10^{34}  B\, g^2_A\, T^6_{10}\,\rho_{14}\,
\biggl(  {A^2   \over  0.1}   \biggr)   \biggl(  {\kappa_0  \over
\mu_\pi  -  \mu_e}  \biggr)^2  \biggl(  {938 MeV \over {m_N^\ast
}}\biggr)\, I\,{\rm ergs \,cm^{-3}\, s^{-1}} &(23)\cr}$$
\vskip .2 cm
\noindent where
$  \kappa_0,\,  (\mu_\pi-\mu_e) $  are in units of MeV,
$\mu_e$ bieng the electron chemical potential.
\vskip .25 cm
The  right  handed  neutrino  pair  emission process
$$  \tilde  n  \longrightarrow   \tilde  n + \nu_R + \bar \nu_R
\eqno{(24)}$$
\noindent is mediated by weak  right handed neutral current which
after chiral rotation becomes
$$  \eqalignno{\tilde  J^{h0}_{R\mu}=&\, {1 \over \cos^2\theta_W}\,
\biggl[ \Bigl(
V_\mu^3  +A_\mu^3   \cos\theta_W  \Bigr)  \cos  \theta  &\cr
&+  \Bigl(
V_\mu^2  \cos  2  \theta_W   +   A_\mu^2  \Bigr)  \sin  \theta  -
\Bigl(  {V_\mu^Y  \over  2 }+ V_\mu^3  \cos \theta + A_\mu^3 \sin
\theta \Bigr) \biggr] &(25)\cr}$$
\vskip .2 cm
\noindent  Unlike   the  charged  current,  here  the momentum is
supplied  only by the wave  function.  The matrix element squared
summed over spins is given by
$$\sum \bigl  \vert   M\bigr \vert^2  =  {{32  \pi^3  G^2_R  g^2_A A^2
\kappa^2_0  \cos^2   2   \theta_W}   \over {\cos^4\theta_W \bigl(
\omega_1  +  \omega_2 \bigr)^2  }}\, \biggl[ \delta^3 (\bar \beta +
\bar  k_0  )  +  \delta^3  (\bar  \beta  -  \bar  k_0  )  \biggr]
 \,\omega_1 \omega_2    \eqno{(26)}$$
\vskip .2 cm
\noindent  and  emissivity  is  calculated  to  be
 $$   \eqalignno   {\dot {\cal E} &=    {{G^2_R   \cos^2
 2\theta_W}
\over   {15   \pi^3   \cos^4   \theta_W}}\,   g^2_A\,  A^2\, \kappa_0^2
\,\mu_\pi^4\, n_N &\cr
&  \simeq  2.5 \times 10^{41} B' \,g^2_A \,\rho_{14} \,\biggl(
{A^2 \over 0.1} \biggr) \biggl( {\mu_\pi  \over 250 MeV} \biggr)^4
\biggl( {938 MeV \over {m_N^\ast}} \biggr) \quad {\rm ergs\, cm^{-3}\, s^{-1}}
&
(27)\cr}$$

\beginsection {3. Neutrino Trapping}

\indent Let   us   now   estimate  the  mean  free  path  of right handed
neutrinos in the  presence  of charged and neutral pion condensed
matter, likely to be present in the core. The important reactions
that  contribute  to the neutrino mean free path are the neutrino
absorption processes
$$ \nu_R +  \eta \longrightarrow \eta + e^- \eqno{(28)}$$
\noindent and
$$  \nu_R + \tilde n \longrightarrow \tilde p + e^- \eqno{(29)}$$
\noindent  In  addition  we  have  the  $\nu_R$  scattering
processes
$$   \nu_R   +   \tilde   N\longrightarrow  \nu_R  +   \tilde   N
\eqno{(30)}$$
\noindent  in  neutral  pion  condensate. The first two processes
are  mediated  by  charged  current  and  the last one by neutral
current.    The    absorption   length  $  \mit  l  \bigl(\omega_\nu,
\,T  \bigr)$ for  a  right  handed  neutrino of
energy  $  \omega_\nu$ in  a medium of temperature  T  is
defined  as $[15]$
$$\eqalignno    {\mit  l^{-1}\bigl(  \omega_\nu,\,  T  \bigr)  =& {1
\over
(2  \pi)^5}  \int  d^3p_1  \int  d^3p_2 \int d^3p_e \sum \bigl\vert  M
\bigr\vert^2\, \delta^4 \bigl(p_1 + q_\nu - k - p_2  -    q_e  \bigr) &\cr
& {\hskip 1 cm}\times \,\Bigl[  1-f  \bigl(  \omega_e,  \mu_e  \bigr)  \Bigr]\,
\Bigl[  1-f
\bigl(  E_2,  \mu_2  \bigr)  \Bigr] \,f\bigl( E_1, \mu_1 \bigr)   &
(31)\cr }$$
\vskip .2 cm
\noindent Using $\sum \bigl\vert M \big\vert^2 $ derived in section 2,
we get
\vskip .25 cm
$$   \eqalignno  {\mit  l^{-1}  &= {G^2_R \over 2 \pi}\, \bigl( 1 +
3 g^2_A  \bigr) \, \Bigl(  1  +  {g^2_A k^2_c \over \mu^2_\pi} \Bigr)\,
\theta^2 \,n_N\, \bigl( \omega_\nu +k_0 \bigr)^2\,
\Bigl[  1+  e^{ \bigl(  \mu_e - \mu_\pi
- \omega_\nu \bigr) /T} \Bigr]^{-1} &\cr
&   \simeq   0.32\,  B\,  \rho_{14} \, \bigl( 1 + 3 g^2_A \bigr)\, \Bigl(
1  +  {g^2_A  k^2_c  \over \mu^2_\pi} \Bigr) \Bigl( {\theta^2
\over   0.1}  \Bigr)&\cr &{\hskip 1 cm}\times \Bigl(  {938  MeV  \over
{m_N^\ast}} \Bigr)
\Bigl[ 1+ e^{  \bigl(  \mu_e - \mu_\pi - \omega_\nu \bigr) /
T} \Bigr]^{-1}\quad {\rm m^{-1}}  &(32) \cr}$$
\vskip .25 cm
$$   \eqalignno   {\mit  l^{-1}  &=  {G^2_R \, A^2\, \kappa_0^2\, g^2_A
\over
2 \,  \pi} \,  n_N \,  \Bigl(  {  \omega_\nu + k_0 \over 2 \omega_\nu +
\mu^0_\pi}    \Bigr)^2   \, \Bigl[   1+   e^{\bigl(   \mu_e   -
\mu^0_\pi  -  \omega_\nu \bigr)/ T} \Bigr]^{-1}&\cr
&  \simeq 1.2\,  B \, \rho_{14}\,  \Bigl(  {A^2  \over  0.1}  \Bigr)  \Bigl(
{\kappa_0 \over 104.57 MeV} \Bigr)^2\, \Bigl( {938 MeV \over {m_N^\ast}}\,
\Bigr)  &\cr
&{\hskip 1 cm}\times\Bigl(  1-  {\omega_\nu  \over   \mu^0_\pi   +   2
\omega_\nu}   \Bigr)   \Bigl[  1+  e^{ \bigl( \mu_e - \mu^0_\pi
-     \omega_\nu     \bigr)/T}    \Bigr]^{-1}\quad {\rm   m^{-1}}
&(33)\cr}$$
\vskip .25 cm
\noindent for  processes  $(28)$  and  $(29)$  respectively.  For
the scattering process, the mean free path is likewise calculated
to be
\vskip .25 cm
$$    \eqalignno   {\mit   l^{-1}  &  =   {4\,G^2_R\,   g^2_A \, A^2
\,\kappa^2_0 \,n_N \, \cos^2  2\theta_W  \over   \pi\,\cos^4\theta_W}\,  \Bigl(
 1+
{\omega_\nu \over  \omega_\nu + \mu_\pi}  \Bigr)^{-2} &\cr
& {\hskip 2 cm}\times \Bigl[  1+
e^{- \bigl( \mu^0_\pi + \omega_\nu \bigr) / T} \Bigr]^{-1}
&\cr
& \simeq 4\, B'\, \rho_{14}\, \Bigl( {A^2 \over 0.1}  \Bigr)\, \Bigl(
{\kappa_0  \over  104.57  MeV}  \Bigr)^2 \, \Bigl( {938  MeV  \over
{m_N^\ast}}  \Bigr)&\cr
& {\hskip 2 cm}\times\Bigl(  1+ {\omega_\nu \over  \mu^0_\pi +
2  \omega_\nu}  \Bigr)^{-2}  \,\Bigl[  1+  e^{- \bigl( \mu^0_\pi +
\omega_\nu \bigr)  T} \Bigr]^{-1} \quad{\rm m^{-1}} &(34)\cr}$$

\beginsection {4. Results and Discussions}

\indent An   estimate   of   the   allowed  range  of  the  right  handed
current    interactions    can   be   made   by   the   following
considerations.  If  the neutrino absorption length happens to be
greater than the core radius, then  $\nu_R$  streams  out  freely
cooling   the   core faster. As $G_R $ increases,  the  flux  for
$\nu_R$  also  increases.  Hence the total integrated  luminosity
along   with   the   constraint  that the total energy available  for
emission   {\it   via}  $\nu_R$  associated with SN 1987 A cannot
exceed  $2-4  \times 10^{53}$ ergs and that $\nu_R$ alone should
not   cool  the core in the time scale less than $5-10\, sec.$, set
an  upper  bound  on  $G_R.$  On  the other hand if the
neutrino mean free  path  is  less  than   the
radius  of  the  core  ,  they
would   then   be  trapped and thermalised  emission  of  $\nu_R$
would  now  proceed  through standard thermal  diffusion  process
from  right  handed  neutrino  sphere.  In this case  as has been
discussed by Barbieri and Mohapatra $[8]$, energy loss {\it  via}
$\nu_R$   will   dominate  unless  $G_R$  satisfies certain lower
limit.   These   constraints  on  the  strength  of  right handed
interaction  translates  into  a  range  of  excluded  values  of
$M_{W_R}$ and $M_{N'}.$

{}From   the  above  consideration  the  upper  limit  on $G_R$ for
charged   current   interactions   is   obtained  from  equations
$(12,23)$   and   for  the  neutral   current   interaction  from
equations  $(16,27)$.  For  the  trapping to  occur  we  see from
equations  $(28-30)$  that   $B  \ge  4 \times 10^{-6}$ and
following   the   reference   $[8]$  we obtain the lower limit on
$G_R.$  These    limits   tranlate  into  an  excluded  range  of
$M_{W_R}$   and  $M_{N'}$ masses. For Charged
current interaction the excluded range is  given  by
    $$   10^{-4} \approx\leq   \biggl[   \xi^2  +  {M^2_{W_L}
    \over  M^2_{W_R}}
\biggr]^{1   \over   2}  \le  \approx 0.1-0.3$$
\noindent  and  for  the  neutral  current interaction
  $$   1.41  \times  10^{-5}\approx \le \biggl[ \xi^2 + {M^2_{W_L}\over
M^2_{N'}}  \biggr]^{1 \over 2} \le \approx 0.1-0.3$$
\noindent  In  the  absence of  $W_L-W_R$  mixing,  the  excluded
range  for  $M_{W_R}$ and   $M_{N'}$   become
 $(1.8-3.1) \,M_{W_L}\,\approx  \le  M_{W_R}  \le\approx\, 100\,
 M_{W_L}  $ and $(1.8-3.16)\, M_{W_L}\,\approx \le  M_{N'}
 \le \,\approx 225 \, M_{W_L}$ We  find
that   in the  presence   of  pion  condensate   in  the core of the
nascent  neutron star the upper bound on the  excluded region for
 $M_{N'}$ is stronger in comparision to that of reference $[8]$.

\beginsection {5. Acknowledgement}

\indent We would like to thank Professor S.R. Choudhury for his encouragement
and
helpful discussion. We also thank Professor J.V. Narlikar for providing
hospitality at Inter-University Centre for Astronomy and Astrophysics, Pune.

\beginsection {6. References}

{\parindent 15 pt
\item{1.}  J. Kogut  and  J.T. Manassah, Phys. Lett. {\bf 41}, 129, (1972);
A.B.  Migdal,   Phys. Rev. Lett.  {\bf  31}, 247, (1973);  R.F. Swayer
and  D.J. Scalpino, Phys. Rev. {\bf D7}, 953, (1973); R.F. Dashen and
J.T. Manassah,  {\it    ibid}, {\bf    12}, 979, (1975);   {\bf  12},
1010, (1975);  G. Baym,   D. Campbell, R. Dashen   and   J. Manassah,
Phys.  Lett. {\bf   B58}, 304, (1975);   D.B. Kaplan   and  A.E. Nelson,
Phys.  Lett. {\bf     B175}, 57, (1986);    {\it    ibid    }, {\bf
B179}, 409E, (1986);   G.E. Brown,  K. Kubodera  and M. Rho, {\it ibid}
{\bf   B192}, 273, (1987);   T. Muto and T. Tatsumi, {\it ibid } {\bf
B283}, 165, (1992);     G.E. Brown,    K. Kubodera,   M. Rho   and
V. Thorrsson, {\it ibid} {\bf B291}, 355, (1992).

\item{2.}    R. Mayle,    D.N. Schramm,    M.S. Turner    and   J.R. Wilson,
Phys. Lett. {\bf       B317}, 19, (1993); B. Friedman,
V.R. Pandharipande  and  Q.N. Usmani, Nucl. Phys. {\bf A372}, 483, (1981).

\item{3.} J.M. Lattimer,          C.J. Pethick,        M. Prakash        and
P. Haensal,            Phys. Rev. Lett. {\bf         66}, 2701, (1991);
M. Prakash, M. Prakash, J.M. Lattimer         and         C.J. Pethick,
Astrophys. J. Lett. {\bf      390}, 77, (1992); D. Page    and
J.H. Applegate, Astrophys. J. Lett. {\bf 394}, 117, (1992).

\item{4.}      G. Byam      and      S. Chin,      Phys. Lett. {\bf    B62},
 241, (1976); E. Witten, Phys. Rev. {\bf     D30}, 272, (1984);
E. Farhi  and  R.L. Jaffe,  Phys. Rev. {\bf D30}, 2379, (1984).

\item{5.} G. Raffelt and D. Seckel, Phys. Rev. Lett. {\bf 60}, 1793, (1988).

\item{6.}   M.S. Turner,   Phys. Rev. {\bf   D45}, 1066, (1992);   A. Burrows,
R. Gandhi  and  M.S. Turner,  Phys. Rev. Lett. {\bf 68}, 3834, (1992);
A. Goyal and S. Dutta, Phys. Rev {\bf D49}, 3910, (1994).

\item{7.}    J.M. Lattimer  and  J. Cooperstein,  Phys. Rev. Lett. {\bf  61},
 23, (1988);  A. Goyal,S. Dutta  and  S.R. Choudhury  , Phys. Lett. {\bf
B346}, 312, (1995).

\item{8.} R. Barbieri and R.N. Mohapatra, Phys. Rev. {\bf D39}, 1229, (1989).

\item{9.}N. Iwamoto,       Phys. Rev. Lett. {\bf       53}, 1198, (1984);
N. Iwamoto,  Phys. Rev. {\bf    D39}, 2120, (1989); M.S. Turner,
Phys. Rep.{\bf 197}, 67, (1990);  G. Raffelt, Phys. Rep. {\bf 198},
1, (1990).

\item{10.}T. Muto,   T. Tatsumi   and   N. Iwamoto,
Phys. Rev. {\bf D50},
6089, (1994).

\item{11.}{\it    e.g.}   see   review,   R.N. Mohapatra  'Quarks,
Leptons  and  Beyond', Proceedings   of  NATO  Advanced
Study    Institute,  Munich, West Germany   (1983),   edited   by
H. Fritzsch {\it et.al} ( NATO ASI, series {\bf B}, Vol. {\bf 122},
(plenum New York, 1985).

\item{12.} J.Ellis   and   K.Olive   Phys. Lett. {\bf B193}, 525, (1987);
M. Turner,  Phys. Rev. Lett.{\bf 60}, 1699, (1988); R. Mayle
{\it  et.al.}, Phys. Lett. {\bf  B203}, 198, (1988);
R. Barbieri    and    R.N. Mohapatra, Phys. Rev. Lett. {\bf 61},
27, (1988);  {\it  e.g.}  see R.N. Mohapatra and P.B. Pal,
'Massive Neutrinos in Physics   and  Astrophysics',  World
Scietific, (1993).

\item{13.} O. Maxwell, G.E. Brown, D.K. Campbell,   R.F. Dashen        and
J.T. Manassah,  Astrophys. J. {\bf216}, 77, (1977).

\item{14.}T. Tatsumi, Prog. Theor. Phys. {\bf         69}, 1137, (1983);
H. Umeda,       K. Nomoto, S. Tsuruta, T. Muto      and      T. Tatsumi,
Astrophys. J. {\bf   431}, 309, (1994); T. Tatsumi, Prog. Theor. Phys.
{\bf 80}, 22, (1988).

\item{15.}R.F. Sawyer and A. Soni, Astrophys. J. {\bf216}, 73, (1977).
}
\bye